\documentclass[12pt]{iopart}
\usepackage{iopams}
\usepackage{graphicx}
\usepackage{color}
\usepackage[colorlinks=true,linkcolor=blue,pagecolor=blue,filecolor=blue,menucolor=blue,urlcolor=blue,citecolor=blue,anchorcolor=blue]{hyperref}

\begin{document}

\title[Quantum metrology in local dissipative noises]{Quantum metrology in local dissipative environments}
\author{Yuan-Sheng Wang$^1$, Chong Chen$^{1,2}$, and Jun-Hong An$^1$}
\address{$^1$ School of Physical Science and Technology, Lanzhou University, Lanzhou 730000, China}
\address{$^2$ Department of Physics, The Chinese University of Hong Kong, Shatin, New Territories, Hong Kong, China}
\ead{anjhong@lzu.edu.cn}
\vspace{10pt}

\begin{abstract}
Quantum metrology allows us to attain a measurement precision that surpasses the classically achievable limit by using quantum characters. The metrology precision is raised from the standard quantum limit (SQL) to the Heisenberg limit (HL) by using entanglement. However, it was reported that the HL returns to the SQL in the presence of local dephasing environments under the long encoding-time condition. We evaluate here the exact impacts of local dissipative environments on quantum metrology, based on the Ramsey interferometer. It is found that the HL is asymptotically recovered under the long encoding-time condition for a finite number of the probe atoms. Our analysis reveals that this is essentially due to the formation of a bound state between each atom and its environment. This provides an avenue for experimentation to implement quantum metrology under practical conditions via engineering of the formation of the system-environment bound state.
\end{abstract}

\submitto{\NJP}
%
\maketitle
%

\section{Introduction}\label{introduction}
Ultrasensitive metrology plays an important role in physics and other science. Using quantum properties such as squeezing \cite{PhysRevLett.71.1355,Giovannetti1330} or entanglement \cite{PhysRevA.54.R4649,PhysRevLett.98.223601,Nagata726,Jones1166,PhysRevLett.112.103604}, quantum metrology permits substantial improvement to the measurement precision of physical quantities, with respect to the used resource, more efficiently than its classical counterpart. Any metrology is constrained by the existence of errors. The error effects can be reduced by repeating the measurement. It is guaranteed by the center limit theorem which states that the average of a large number $N$ of independent measurements each having a standard deviation $\Delta \sigma$ converges to a Gaussian distribution with standard deviation or precision $\Delta \sigma/\sqrt{N}$ \cite{Giovannetti2011}. It has been found that the metrology precision with quantum probes made up of $n$ entangled particles attains the so-called Heisenberg limit (HL) scaling $n^{-1}$, which surpasses the standard quantum limit (SQL) or shot noise limit scaling $n^{-1/2}$ when uncorrelated particles are used \cite{PhysRevA.47.3554}. Such dramatic enhancement endows quantum metrology \cite{PhysRevLett.96.010401,Giovannetti2011} extensive applications in gravitational wave detection \cite{Schnabel2010,Abadie2011,RevModPhys.86.121}, atomic clocks \cite{PhysRevA.54.R4649,PhysRevLett.82.2207,Leibfried1476,Roos2006,RevModPhys.83.331,RevModPhys.87.637}, quantum imaging \cite{RevModPhys.71.1539,1464-4266-4-3-372,Morris2015,PhysRevA.94.032301,PhysRevX.6.031033}, and even quantum biology \cite{doi:10.1063/1.4724105,PhysRevX.4.011017,Taylor20161}.

However, the practical realization of quantum metrology is limited by the ubiquitous decoherence effect induced by the influence of outer environments \cite{PhysRevLett.79.3865,Banaszek2009,PhysRevA.81.033819,PhysRevA.84.012103,PhysRevLett.109.233601,PhysRevLett.111.120401,PhysRevLett.111.090801,PhysRevA.90.033846,PhysRevA.92.010102,
PhysRevA.91.052105,PhysRevLett.116.120801,Huanglee215}, which causes entanglement degradation to quantum systems. This inspires the emergence of open-system quantum metrology \cite{PhysRevLett.112.120405,1367-2630-16-1-015023}. It was found that the HL returns to the SQL when the involved systems are exposed to local Markovian dephasing environments \cite{PhysRevLett.79.3865}. Further studies revealed that the precision scales as the so-called Zeno limit $n^{-3/4}$ when the non-Markovian effect of the local dephasing environments is considered \cite{PhysRevA.84.012103,PhysRevLett.109.233601,PhysRevA.92.010102}. This has been proven to be universally valid to the precision estimated under short encoding-time condition in the nonsemigroup dynamics induced by all phase-covariant uncorrelated environments \cite{PhysRevLett.116.120801}. Although it was claimed that the HL could be restored, certain additional control strategies, such as decoherence-free-state \cite{PhysRevA.93.062122} encoding in spatially correlated environment \cite{1367-2630-16-7-073039}, adaptive measurements \cite{PhysRevLett.111.090801}, dynamical decoupling \cite{PhysRevA.87.032102,1367-2630-18-7-073034}, and error correction \cite{PhysRevLett.112.080801,PhysRevLett.116.230502,Lu2015NC}, were applied on the system. A natural question is: Can entanglement perform better than achieving Zeno limit in open-system quantum metrology?

Going beyond the widely studied dephasing environment case \cite{PhysRevA.84.012103,PhysRevLett.109.233601,PhysRevA.92.010102}, we study the impacts of local dissipative environments on the Ramsey-spectroscopy-based quantum metrology to estimate the atomic frequency. The dissipative environment, which causes the atomic spontaneous emission, is the main decoherence resource of atom system. We find that the non-Markovian effect play a constructive role in improving the metrology precision in this situation. A novel scaling relation to the metrology precision of the atomic frequency is derived. It is interesting to find that the scaling in the presence of the dissipative environments can approach asymptotically the HL in ideal case even without any additional control strategy applied to the system. Our analysis reveals that the underlying mechanism is the formation of a bound state of the local system consisting of each atom and its environment. This implies that the quantum metrology in the dissipative environment could even perform better than that in the dephasing environment \cite{PhysRevA.84.012103,PhysRevLett.109.233601,PhysRevA.92.010102} in surpassing the SQL. By enriching our understanding of the decoherence mechanism in quantum metrology, our result could be important in the practical realization of quantum metrology via bound-state engineering.

\section{Ideal quantum metrology}
To estimate a parameter $x$ of a physical system, we first prepare a probe and couple it to the system to encode the parameter information. Then we measure the probe and infer the value of $x$ from the results. The inevitable errors make us unable to estimate $x$ exactly. Although they can be reduced by repeating the measurement, the best precision (denoted by the root mean square) in estimating $x$ subject to projective measurements $\sum_i \hat{M}^\dag_i\hat{M}_i=1$ on the probe state $\rho(x)$ is given by Cram\'{e}r-Rao bound \cite{PhysRevA.87.032102,PhysRevA.92.022106,PhysRevA.90.062130,Liu2015scirep}
\begin{equation}
\delta x=[NF(x)]^{-1/2},\label{CRB}
\end{equation}
where $N$ is the measurement times and $F(x)=\sum_{i}[\partial_x p_i(x)]^{2}/p_i(x)$ with $p_i(x)=\mathtt{Tr}[\hat{M}_i\rho(x)\hat{M}^\dag_i]$ is the Fisher information characterizing the most information for estimating $x$ extractable from the measurements. Note that for the explicit metrology scheme using definite type of initial entanglement and measurement method, it generally suffices to use Fisher information to demonstrate the improvement of metrology precision \cite{PhysRevLett.109.233601}. However, for general situation, the quantum Fisher information, which involves the optimization to the measurements, is needed  \cite{Escher2011,demkowicz2012elusive}.

We consider a concrete setup to estimate atomic frequency $\omega_0$ in the Ramsey interferometer \cite{PhysRevLett.79.3865,PhysRevLett.109.233601}. The probe has $n$ atoms as the physical resource. Three steps, i.e., the state preparation, the parameter encoding, and the readout, form the basic procedure. In the case of the uncorrelated input state, the preparation step produces a state $|\Psi_\mathtt{in}\rangle=[(|g\rangle+|e\rangle)/\sqrt{2}]^{\otimes n}$, where $|g\rangle$ and $|e\rangle$ are the atomic ground and excited states. It evolves to $|\Psi_{\omega_0}\rangle=[(|g\rangle+e^{-{\rm i}\omega_0t}|e\rangle)/\sqrt{2}]^{\otimes n}$ in the step of the parameter encoding realized by atomic free evolution. The readout is achieved by checking whether each atom is still in $|\Psi_\mathtt{in}\rangle$, which gives $n$ independent results of ``yes" or ``no" with probability $p_1=\cos^2(\omega_0t/2)$ or $p_2=\sin^2(\omega_0t/2)$. Then $F(\omega_0)=t^2$ can be calculated. Repeating the experiment in time duration $T$, we get $N=nT/t$ data. Then Eq. (\ref{CRB}) induces
\begin{equation}
\delta \omega_0|^\mathtt{ideal}_\mathtt{uncor}=(nTt)^{-1/2}. \label{SQL}
\end{equation}
Such precision scaling to the number $n$ of the physical resource is called SQL \cite{PhysRevA.47.3554}. In the case of the entangled input state, the first step generates a GHZ state $|\Psi_\mathtt{in}\rangle=(|g\rangle^{\otimes n}+|e\rangle^{\otimes n})/\sqrt{2}$, which evolves to $|\Psi_{\omega_0}\rangle=(|g\rangle^{\otimes n}+e^{-{\rm i}n\omega_0t}|e\rangle^{\otimes n})/\sqrt{2}$ in the second step. After performing a CNOT gate with the first atom as the controller and the others as the target, to disentangle them and a Hadamard gate on the first atom, the measuring on the first atom results in $|e\rangle$ with probability $p_1=\cos^2(n\omega_0t/2)$ and $|g\rangle$ with $p_2=\sin^2(n\omega_0t/2)$ in the final step. It leads to $F(\omega_0)=n^2 t^2$. Repeating the experiment in duration $T$, we have $N=T/t$ results. Then we have
\begin{equation}
\delta \omega_0|^\mathtt{ideal}_\mathtt{ent}=(n^2Tt)^{-1/2},\label{HL}
\end{equation}
which is called HL. Obviously, the HL (\ref{HL}) using entanglement has a $n^{1/2}$ time enhancement over the SQL (\ref{SQL}).

\section{Effects of dissipative environments} \label{effects}
In practice, the atomic free evolution in the parameter encoding step is inevitably obscured by the presence of the environment. Determined by whether a system has energy exchange with the environment, the impact of the environment on system can be classified physically into dissipation and dephasing. The dephasing arises from elastic collisions in dense atomic ensemble or elastic phonon scattering in a solid system \cite{Carmichael}, neither of which are significant in our countable atom scenario. Thus, in contrast to Refs. \cite{PhysRevLett.79.3865,PhysRevA.84.012103,PhysRevLett.109.233601,PhysRevA.92.010102}, where the dephasing environments have been studies, we here focus on the action of dissipative environments on quantum metrology.

Consider each atom to be subjected to local dissipative environments in the parameter encoding step. The Hamiltonian of the $j$th atom and its environment is $\hat{H}_j=\hat{H}_{j,S}+\hat{H}_{j,E}+\hat{H}_{j,I}$ with $\hat{H}_{j,S}=\omega_0\hat{\sigma}_j^+\hat{\sigma}_j^-$, $\hat{H}_{j,E}=\sum_{k}\omega_k\hat{a}_{j,k}^\dag\hat{a}_{j,k}$, and $\hat{H}_{j,I}=\sum_{k}g_k(\hat{a}_{j,k}\hat{\sigma}^+_j+\mathtt{H.c.})$. Here $\hat{\sigma}_j^\pm$ is the transition operator of the $j$th atom, $\hat{a}_{j,k}$ is the annihilation operator of the $k$th mode with frequency $\omega_k$ of the environment felt by the $j$th atom. The atom-environment coupling strength is $g_{k}=\omega_0\mathbf{\hat{e}} _{k}\cdot\mathbf{d}/\sqrt{2\varepsilon _{0}\omega _{k}V}$, where $\mathbf{\hat{e}}_{k} $ and $V$ are the unit polarization vector and the normalization volume of the environment, $\mathbf{d}$ is the atomic dipole, and $\varepsilon _{0}$ is the free space permittivity. After tracing the $n$ independent environments, the exact master equation governing the parameter encoding reads \cite{breuer2002theory}
\begin{equation}
 \dot{\rho}(t)=\sum_{j=1}^{n}\{-{\rm i}{\omega(t)\over2}[\hat{\sigma}_{j}^{+}\hat{\sigma}_{j}^{-},\rho(t)]+{\gamma(t)\over 2}\check{\mathcal{L}}_{j}\rho(t)\} ,\label{master}
\end{equation}
where $\check{\mathcal{L}}_j\cdot=2\hat{\sigma}_j^-\cdot\hat{\sigma}_j^+-\{\cdot,\hat{\sigma}_j^+\hat{\sigma}_j^-\}$ and $\gamma(t)+{\rm i}\omega(t)=-2\dot{c}(t)/c(t)$ with $c(t)$ determined by
\begin{equation}
  \dot{c}(t)+{\rm i}\omega_{0}c(t)+\int_{0}^{t}f(t-\tau)c(\tau)d\tau=0 \label{int-diff}
\end{equation}
under the condition $c(0)=1$. Here $f(t-\tau)\equiv \int J(\omega){\rm e}^{-{\rm i}\omega(t-\tau)}d\omega$ is the environmental correlation function with $J(\omega)=\sum_{k}|g_{k}|^{2}\delta(\omega-\omega_{k})$ being the spectral density. All the non-Markovian effect characterized by the convolution in Eq. (\ref{int-diff}) has been self-consistently incorporated into the time-dependent renormalized frequency $\omega(t)$ and the decay rate $\gamma(t)$ in Eq. (\ref{master}). The solution of the master equation (\ref{master}) can be generally written as the Kraus representation $\rho(t)=\Lambda^{\otimes n}_{\omega_0}\rho(0)$ with $\Lambda_{\omega_0}\cdot=\sum_i\hat{K}_i(\omega_0)\cdot\hat{K}_i^\dag(\omega_0)$ and
\begin{equation}\label{krs}
\hat{K}_0(\omega_0)=\left(
                      \begin{array}{cc}
                        c(t) & 0 \\
                        0 & 1 \\
                      \end{array}
                    \right),~\hat{K}_1(\omega_0)=\left(
                      \begin{array}{cc}
                        0 & 0 \\
                        \sqrt{1-|c(t)|^2} & 0 \\
                      \end{array}
                    \right)
\end{equation}
represented on the basis defined by $|g\rangle$ and $|e\rangle$. It is a non-full-rank quantum channel, to which the so-called ``classical simulation" approach is not applicable \cite{demkowicz2012elusive}. Although the less-intuitive ``channel extension" approach reveals that such channel under the Markovian approximation is limited by the SQL \cite{Escher2011,demkowicz2012elusive}, we here show that the SQL is surpassable for the considered non-full-rank quantum channel in the non-Markovian case.

The evolution of the two initial states can be calculated from the Kraus representation as \cite{breuer2002theory}
\begin{eqnarray}
\rho(t)|_\mathtt{uncor}=\{{1\over 2}[(2-|c(t)|^2)|g\rangle\langle g|+|c(t)|^2|e\rangle\langle e|\nonumber \\
+c(t)|e\rangle\langle g|+c^*(t)|e\rangle\langle g|]\}^{\otimes n},\\
\rho(t)|_\mathtt{ent}={1\over 2}\{(|g\rangle\langle g|)^{\otimes n}+[c(t)|e\rangle\langle g|]^{\otimes n}+[c^*(t)|g\rangle\langle e|]^{\otimes n}\nonumber\\
+[|c(t)|^2|e\rangle\langle e|+(1-|c(t)|^2)|g\rangle\langle g|]^{\otimes n}\}.
\end{eqnarray}
Then repeating the same readout process as for the ideal case, we can evaluate
\begin{eqnarray}
\delta \omega_0|_\mathtt{uncor} &=&\left\{{nT[\partial _{\omega_0}\mathtt{Re}(c(t))]^2\over t[ 1-\mathtt{Re}^2(c(t))] }\right\}^{-1/2}, \label{omegaduc}\\
\delta \omega_0|_\mathtt{ent}&=&\left\{{T[\partial _{\omega_0}\mathtt{Re}(c^n(t))]^2 \over t[ 1-\mathtt{Re}^2(c^n(t))]}\right\}^{-1/2}.\label{omegadent}
\end{eqnarray}
In the ideal case, $\lim_{g_k\rightarrow0}c(t)=\exp(-{\rm i}\omega_0t)$ and thus Eqs. (\ref{omegaduc}) and (\ref{omegadent}) return to Eqs. (\ref{SQL}) and (\ref{HL}), respectively.

When the atom-environment coupling is weak and the characteristic time scale of $f(t-\tau)$ is much smaller than the one of the atom, we can apply Markovian approximation to Eq. (\ref{int-diff}) and obtain $c(t)=\exp[-(\tilde{\gamma}/2+{\rm i}(\omega_0+\Delta\omega))t]$ with $\tilde{\gamma}=2\pi J(\omega_0)$ and $\Delta\omega=\mathcal{P}\int{J(\omega)\over\omega-\omega_0}d\omega$. Substituting $c(t)$ in Eqs. (\ref{omegaduc}) and (\ref{omegadent}) and optimizing $\omega_0$ and $t$, we obtain $\min(\delta\omega_0|_\mathtt{uncor})=\min(\delta\omega_0|_\mathtt{ent})=(nT/\tilde{\gamma} {\rm e})^{-1/2}$ when $\omega_0 t=k\pi/2$ and $k\pi/(2n)$ for odd $k$ and $t=1/\tilde{\gamma}$ and $1/(n\tilde{\gamma})$, respectively. Thus the metrology precision using the GHZ-type entanglement becomes the exactly same as the one using the uncorrelated input. It means that the advantage of using the GHZ-type entanglement in quantum metrology entirely disappears under the Markovian dissipative environments. Note that this result is obtained without optimizing to the measurements and thus cannot be directly seen as the best metrology precision under the dissipation. More general discussion resorting to measurement optimization via quantum Fisher information has shown that the best metrology precision in the Markovian spontaneous emission quantum channel is limited by the SQL \cite{Escher2011,demkowicz2012elusive}, with which our result is consistent.

In the non-Markovian dynamics, we can solve Eq. (\ref{int-diff}) using Laplace transform and obtain $\tilde{c}(s)=[ s+{\rm i}\omega_0+\tilde{\mu}(s)]^{-1}$ with $\tilde{\mu}(s)=\int_{\omega_c}^\infty{J(\omega)\over s+{\rm i}\omega}d\omega$ and $\omega_c$ being the bottom of the environmental energyband. According to Cauchy residue theorem, the inverse Laplace transform of $\tilde{c}(s)$ can be done by finding the poles of $\tilde{c}(s)$ via
\begin{equation}
y(E)\equiv\omega_0+\int_{\omega_c}^\infty {J(\omega)\over E-\omega}d\omega=E,~(E={\rm i}s).\label{yee}
\end{equation}
Note that the roots of Eq. (\ref{yee}) are just the eigenenergies in single-excitation subspace of the whole system consisting of each atom and its environment \cite{PhysRevA.81.052330}. To see this, we expand the eigenstates as $|\Phi_1\rangle=u_0|e,\{0_k\}\rangle+\sum_k v_k|g,1_k\rangle$. Then from the eigenequation $\hat{H}_j|\Phi_1\rangle=E|\Phi_1\rangle$, we obtain $(E-\omega_0)u_0=\sum_k g_kv_k$ and $v_k=g_ku_0/(E-\omega_k)$, which leads to Eq. (\ref{yee}) readily in the continuous limit of the environmental frequency. It is understandable based on the fact that the atomic dissipation induced by the vacuum environment is dominated by the single-excitation process of the whole system. Since $y(E)$ is a monotonically decreasing function when $E<\omega_c$, Eq. (\ref{yee}) has one discrete root if $y(\omega_c)<\omega_c$. It has an infinite number of roots in the region $E>\omega_c$, which form a continuous energyband. We name this discrete eigenstate with the eigenenergy $E_0<\omega_c$ bound state. Its formation would have profound consequences on the atomic dynamics. To see this, we take the inverse Laplace transform and obtain
\begin{equation}
c(t)=Ze^{-{\rm i}E_0t}+\int_{{\rm i}\epsilon+\omega_c}^{{\rm i}\epsilon+\infty}{dE\over 2\pi}\tilde{c}(-{\rm i}E){\rm e}^{-{\rm i}Et},\label{invlpc}
\end{equation}
where the first term with $Z=[1+\int_{\omega_c}^\infty{J(\omega)\over (E_0-\omega)^2}d\omega]^{-1}$ is contributed from the potentially formed bound state and the second term contains the contributions from the continuous energyband. Oscillating with time in continuously changing frequencies, the second term in Eq. (\ref{invlpc}) behaves as a decay and approaches zero in the long-time limit due to the out-of-phase interference. Therefore, if the bound state is absent, then $\lim_{t\rightarrow \infty}c(t)=0$, which characterizes a complete decoherence; while if the bound state is formed, then $\lim_{t\rightarrow \infty}c(t)=Z{\rm e}^{-{\rm i}E_0t}$, which implies that the atomic dissipation is inhibited and the entanglement can be partially preserved in the steady state \cite{PhysRevA.81.052330}.

\begin{figure}
  \centering
  \includegraphics[width=.7\columnwidth]{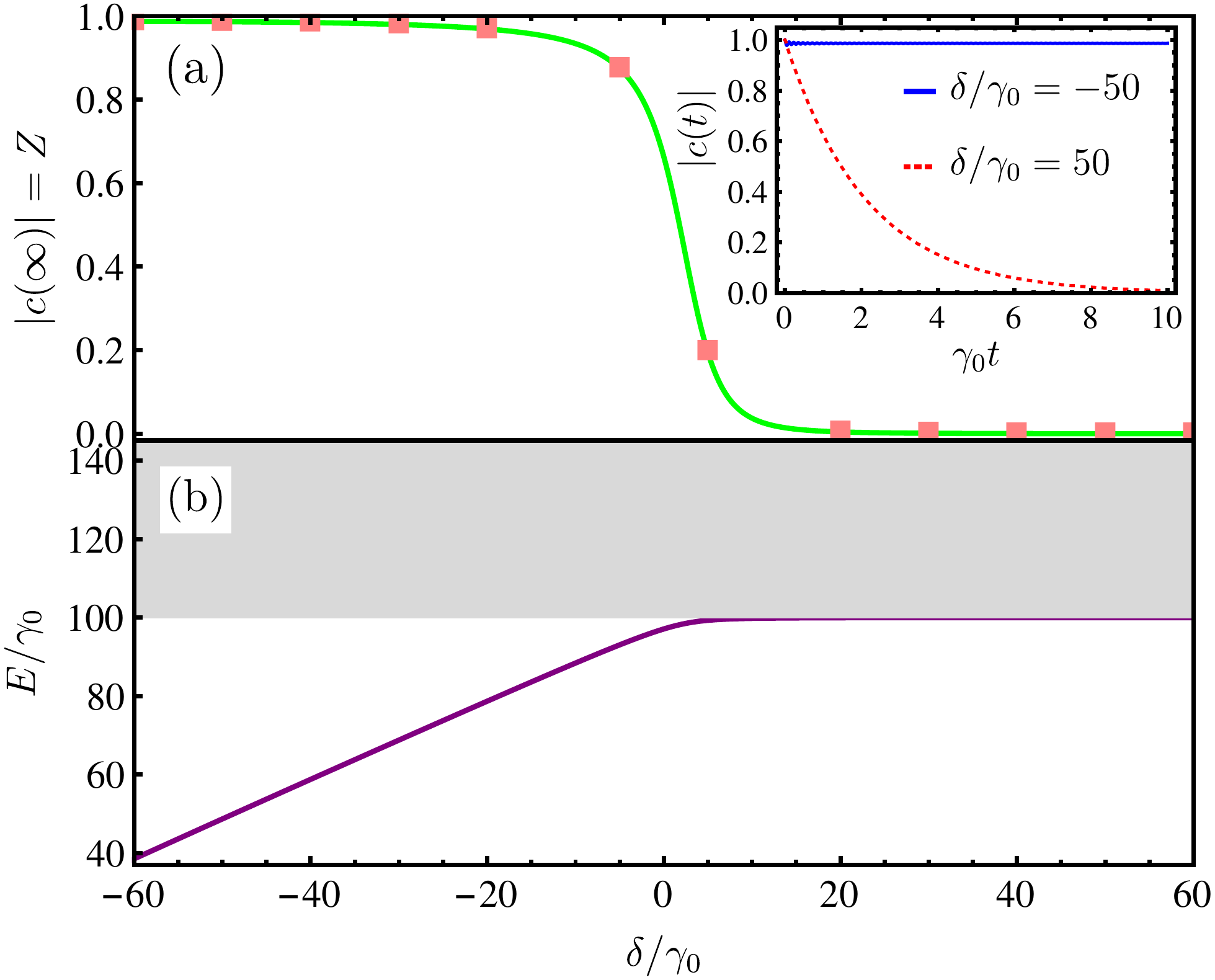}\\
  \caption {(a): Steady-state $|c(t)|$ in different detuning $\delta$ evaluated from Eq. (\ref{amplitude}) (marked by squares), and from $Z$ (green solid line). The inset shows the time evolution of $c(t)$, which approaches $Z$ in the long-time limit. (b): Energy spectrum of the atom-environment system, where the gray area denotes the energyband. Here $\omega_c=100\gamma_0$ and $n=10$ are used.}
  \label{ramsey-spec}
\end{figure}
The formation of the bound state significantly affects the attainable metrology precision in the non-Markovian dissipation dynamics. In contrast to Refs. \cite{PhysRevA.84.012103,PhysRevLett.109.233601,PhysRevA.92.010102,PhysRevLett.116.120801}, which focused on the short encoding-time behavior in the dynamics, we concentrate here on the long encoding-time condition. By substituting the form $Ze^{-{\rm i}E_0t}$ of large-time $c(t)$ into Eq. (\ref{omegadent}), we have
\begin{equation}
\min(\lim_{t\rightarrow\infty}\delta \omega_0|_\mathtt{ent})\leq Z^{-(n+1)}(n^2Tt)^{-1/2},\label{bdsprc}
\end{equation}
where the dependence of $E_0$ on $\omega_0$ has been considered via $\partial_{\omega_0}E_0=Z$. Equation (\ref{bdsprc}) would approach the HL (\ref{HL}) for $n\ll \lfloor -1/\ln Z\rfloor$ when $Z$ reaches 1. Representing the contribution of the bound state to the residual excited-state population, $Z^2$ can be controlled by manipulating the environmental spectral density $J(\omega)$. It can be realized by fabricating the spatial confinement of the radiation field such that the dispersion relation of the atomic radiation field is efficiently changed. Such spatial fabrication includes the three-dimensional periodic structure in the photonic crystal, which results in the band-gapped dispersion relation \cite{PhysRevLett.64.2418,PhysRevA.50.1764}, and the two-dimensional quantum surface plasmonics, which results in the strongly surface-confined propagation of the radiation field \cite{Tame2013,PhysRevB.95.161408}. All these exotic environmental characteristics in turn would make the properties of the atomic dissipation changeable. These progresses give us sufficient room to control the atomic dissipation via reservoir engineering. It is remarkable to find that the HL is asymptotically achievable even under this large encoding-time condition.

\section{Physical realization}
To illustrate the significant impacts of the bound state and the non-Markovian effect on the metrology precision, we consider a band-gapped environment in photonic crystal setting. The dispersion relation of such structured environment reads $\omega_{k}=\omega_{c}+A(k-k_{0})^{2}$, where $\omega_{c}$ is the environmental band edge frequency, and $A=\omega_{c}/k_{0}^{2}$ with $k_{0}\simeq\omega_{c}/c$ \cite{PhysRevLett.64.2418,PhysRevA.50.1764,PhysRevA.87.022312,Liu2016}. Using Laplace-transform method, we can calculate
\begin{equation}
c(t)=e^{-{\rm i}\omega_{0}t}\sum_{j=1}^{3}a_{j}{\rm e}^{(i\delta+\beta x_{j}^{2})t}\left[x_{j}+\sqrt{x_{j}^{2}} \mathtt{Erf}(\sqrt{\beta x_{j}^2}t)\right],\label{amplitude}
\end{equation}
where $\beta=\omega_c(\pi\gamma_0/2\omega_0)^{2/3}$ with $\gamma_0=\omega_0^3d^2/(3\pi\epsilon_0c^3)$, $\delta=\omega_{0}-\omega_{c}$, and $a_j=x_j/[(x_j-x_i)(x_j-x_k)]$ ($i\neq j\neq k=1,2,3$) with $x_{j}$ being the solutions of the equation $(\beta x^{2}+{\rm i}\delta)\sqrt{\beta}x-({\rm i}\beta)^{3/2}=0$. Here $\gamma_0$ is the vacuum spontaneous emission rate characterizing the intrinsic lifetime of the atoms. Taking $\gamma_0$ as the frequency scale, we will neglect its dependence on $\omega_0$. We can analytically prove
\begin{equation}
\lim_{|\delta|\rightarrow\infty}|c(t)|=\big|[1+{1\over 2}(-{\beta/ \delta})^{3/2}]^{-1}e^{-({\beta^3/ \delta})^{1/2}t}\big|.\label{ldet}
\end{equation}
For $\delta>0$, Eq. (\ref{ldet}) decays to zero in constant rate $({\beta^3/ \delta})^{1/2}$. From Eq. (\ref{omegadent}), we have $\lim_{\delta\rightarrow +\infty}\min(\delta\omega_0|_\mathtt{ent})=(nT/\tilde{\gamma} e)^{-1/2}$ with $\tilde{\gamma}=2(\beta^3/\delta)^{1/2}$, which is the SQL and recovers the Markovian result \cite{Escher2011,demkowicz2012elusive}. For $\delta<0$, Eq. (\ref{ldet}) tends to $[1+{1\over 2}(-{\beta/ \delta})^{3/2}]^{-1}$, which matches with our bound-state analysis.
\begin{figure}
  \centering
  \includegraphics[width=.7\columnwidth]{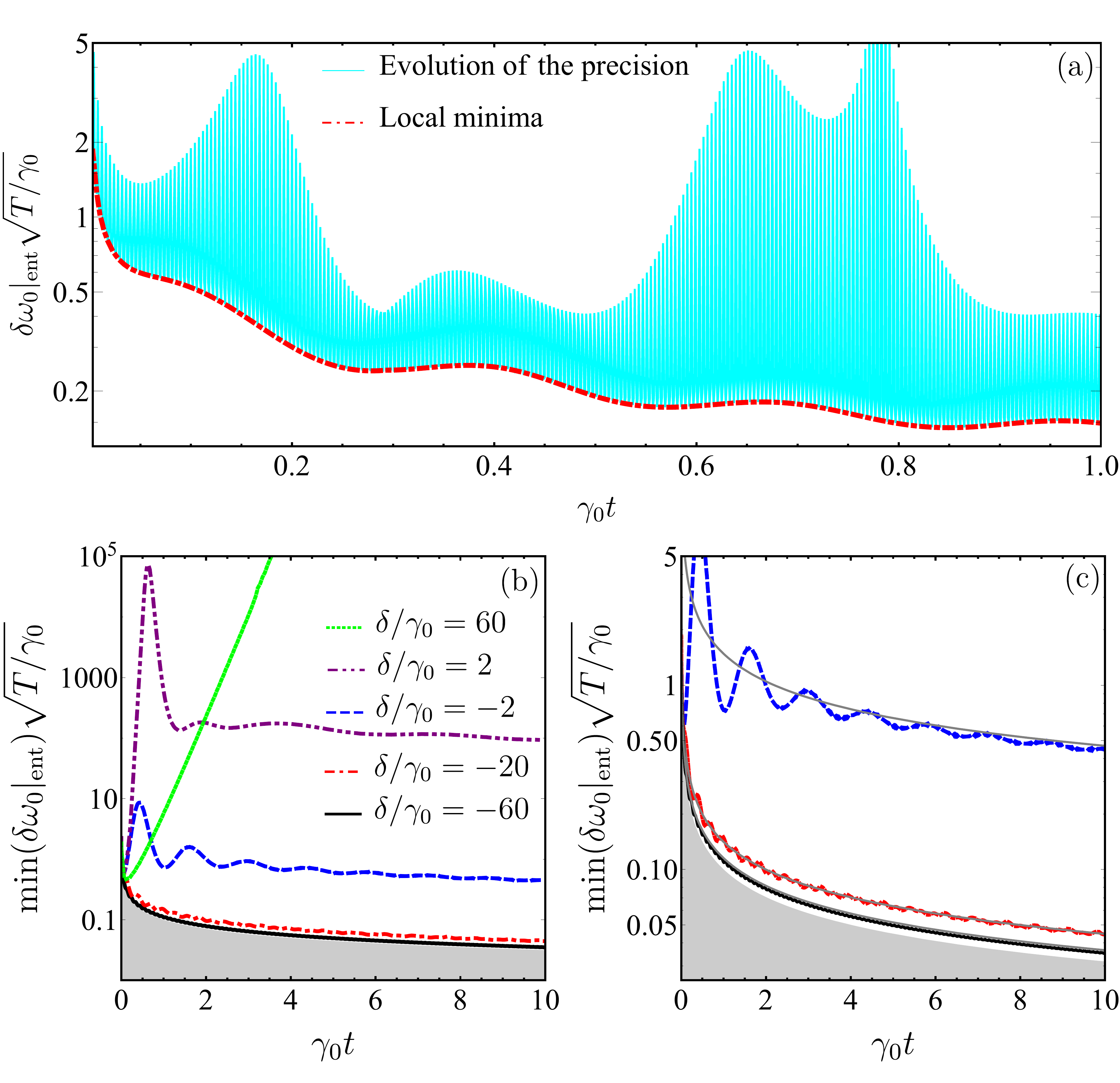}\\
  \caption{(a): Evolution of the precision $\delta\omega_0|_\mathtt{ent}$ (cyan solid line) and the envelope of its local minima (red dot-dashed line) when the detuning $\delta=-20\gamma_0$. (b) and (c): $\mathtt{min}(\delta\omega_0|_\mathtt{ent})$ evaluated from the local minima as function of encoding time $t$ in different $\delta$. The gray area denotes the regime beyond the HL. The gray lines in (c) represent the results evaluated from Eq. (\ref{bdsprc}). Other parameters are the same as Fig. \ref{ramsey-spec}.}
  \label{metrol}
\end{figure}

According to the recent circuit QED experiment \cite{Liu2016}, the band edge frequency $\omega_c\simeq8.0$ GHz, the vacuum spontaneous emission rate $\gamma_0\simeq 50$ MHz, and the atomic bare frequency $\omega_0\simeq 6.0$ GHz to 8.5 GHz. Using these parameters, we perform the numerical calculation. Figure \ref{ramsey-spec}(a) shows the numerical result of long-time $|c(t)|$ in different $\delta$. One can check from Eq. (\ref{master}) that $|c(t)|^2$ just represents the time-dependent factor of the excited-state population of each atom. It is interesting to find that, when $\omega_0<\omega_c$, the atom can be partially stabilized in its excited state even in the presence of the dissipative environment. This is quite different from the Markovian approximation and the bound-state-absent results where the atom decays completely to its ground state. Figure \ref{ramsey-spec}(b) is the energy spectrum of the whole system formed by each atom and its environment. It shows that the regime where the atomic excited-state population is preserved matches well with the one where a system-environment bound state is formed in the environmental bandgap. It is physically understandable based on the fact that the bound state, as a stationary state of the whole system, would preserve the excited-state population in its superposed components during time evolution \cite{PhysRevA.87.022312}. Thus $|c(t)|$ approaches the value $Z$ contributed uniquely from the bound state [see Fig. \ref{ramsey-spec}(a) and its inset]. It is remarkable to find that $|c(t)|$ and $Z$ even tend to 1 when the atomic frequency falls in the deep of the environmental energy bandgap.

\begin{figure}
  \centering
  \includegraphics[width=.7\columnwidth]{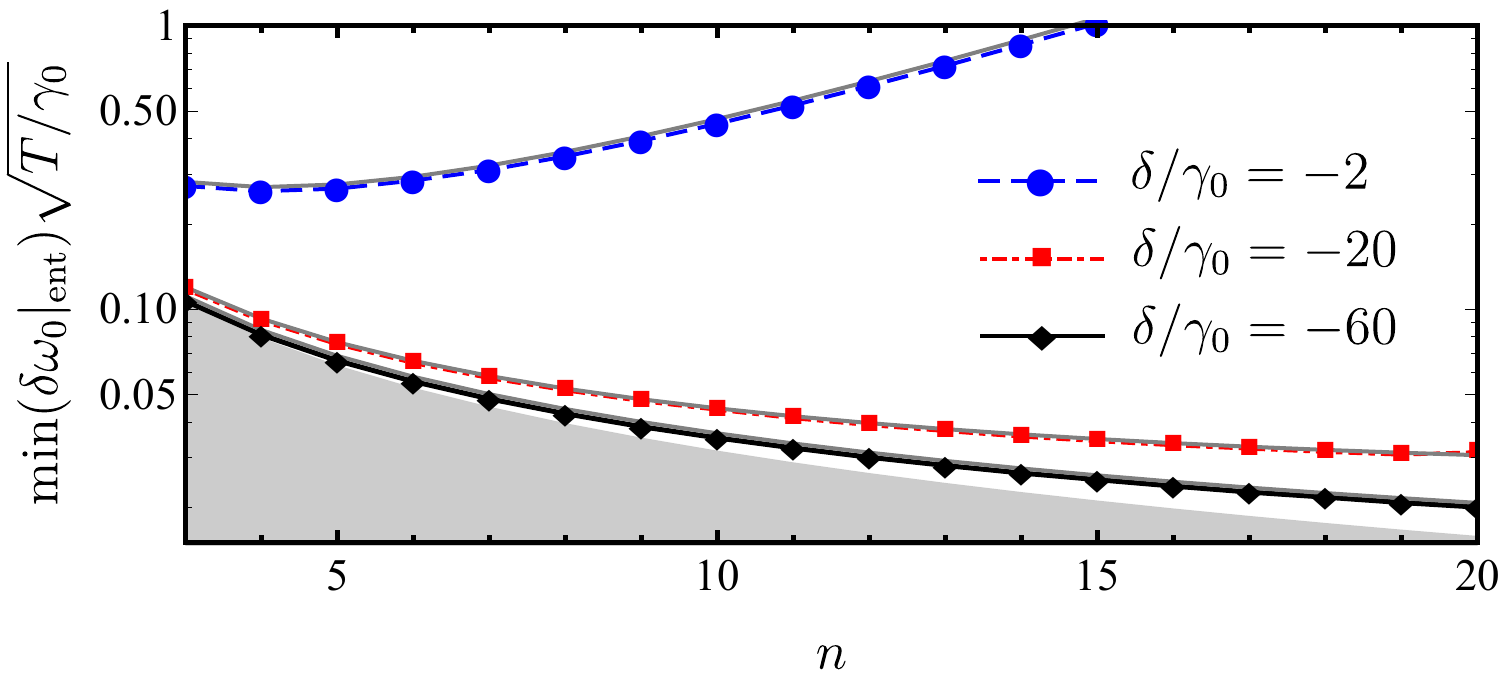}\\
  \caption{Minimal precision of $\delta\omega_0|_\mathtt{ent}$ as function of the atomic number $n$ at different $\delta$. The encoding time is chosen as $t=10/\gamma_0$. The gray area denotes the regime beyond the HL. The gray lines represent the results evaluated from Eq. (\ref{bdsprc}). Other parameters are the same as Fig. \ref{ramsey-spec}.}
  \label{HLM2}
\end{figure}
With the analytical form (\ref{amplitude}) at hand, we can readily obtain the metrology precision $\delta\omega_0|_\mathtt{ent}$ from Eq. (\ref{omegadent}). Figure \ref{metrol}(a) shows time evolution of $\delta\omega_0|_\mathtt{ent}$ as function of the encoding time $t$ for definite detuning $\delta$. Oscillating with time, $\delta\omega_0|_\mathtt{ent}$ takes its best values as the envelope of the local minima [see red dot-dashed line in Fig. \ref{metrol}(a)]. In this way, we can evaluate the best metrology precision $\min(\delta\omega_0|_\mathtt{ent})$ in different detuning $\delta$ [see Fig. \ref{metrol}(b)]. One can find that with the formation of the bound state in negative $\delta$ regime, $\min(\delta\omega_0|_\mathtt{ent})$ decreases with increasing $t$. It is in sharp contrast to the case without the bound state in positive $\delta$ regime, where it increases and the metrology gets worse with increasing $t$ \cite{PhysRevLett.79.3865}. It is surprising to find that on further decreasing $\delta$ in the bound-state regime, $\min(\delta\omega_0|_\mathtt{ent})$ can even reach the HL. It reveals the significant effect induced by the bound state in noisy quantum metrology. To further verify this result, we compare in \ref{metrol}(c) the precision with the scaling (\ref{bdsprc}). After shot-time jolting, the precision tends to the scaling (\ref{bdsprc}). Setting the encoding time as large as $t\simeq10/\gamma_0$, we plot in Fig. \ref{HLM2} $\min(\delta\omega_0|_\mathtt{ent})$ as function of atomic number $n$. Once again, it verifies the validity of the scaling (\ref{bdsprc}). Furthermore, with decreasing $\delta$, the precision gets nearer and nearer the HL. It confirms our expectation that the precision asymptotically matching the analytical scaling (\ref{bdsprc}) approaches the HL for very small $\delta$. It implies that, due to the distinguished role played by the bound state, quantum metrology under independent dissipation can perform even better than the one under independent dephasing, where the precision scales as the Zeno limit $n^{-3/4}$ for the short encoding time and the SQL for the large encoding time \cite{PhysRevA.84.012103,PhysRevLett.109.233601,PhysRevA.92.010102}. Note that, due to the competition between the prefactor $Z^{-(n+1)}$ and the HL in Eq. (\ref{bdsprc}) as function of $n$, the HL can only be approached for $n\ll\lfloor -1/\ln Z\rfloor$, which goes to infinite for $Z$ tending to $1$. Therefore, there is a balance in enhancing precision, between increasing $n$ and dissipation factor $Z$.

Our result is confirmable in the circuit QED platform \cite{Liu2016}, where the bound state has been observed in photonic band-gapped environment. Note that our result is extendable to other spectral densities \cite{PhysRevA.81.052330,PhysRevB.95.161408}, where the formation mechanism of the bound state is the same. Although only the dissipative quantum metrology based on Ramsey interferometer is studied, the revealed mechanism is applicable to both the magnetic field sensing \cite{PhysRevA.84.012103} and Mach-Zehnder interferometer \cite{PhysRevLett.107.083601}, where the dissipation of the quantized light has the same dynamical equation as Eq. (\ref{int-diff}) \cite{1751-8121-42-1-015302}.

\section{Conclusions}\label{conclusion}
In summary, we have investigated quantum metrology based on Ramsey spectroscopy, in the presence of local non-Markovian dissipative environments. It has been revealed that, the metrology precision in dissipative environments can asymptotically approach the HL in long encoding-time condition for finite number of probe atoms, which is quite different from the pure dephasing environment case. This discovery is attributed to the formation of a bound state between each atom and its quantum environment. Our result suggests an active way to realize the ultrahighly precise measurement in the practical dissipation-environment situation by engineering the formation of the bound state \cite{Liu2016}. It could play instructive role in frequency estimation experiments and be generalized readily to other estimation scenarios.

\ack
This work is supported by the National Natural Science Foundation (Grant No. 11474139) and by the Fundamental Research Funds for the Central Universities of China.

\section*{References}
\bibliographystyle{iopart-num}
\bibliography{BibLib}
\end{document}